\documentclass[12pt]{iopart}
\usepackage{iopams}
\usepackage{epsfig}

\begin{document}

\title[]{Hyperon production in Pb+Pb collisions \\ at the CERN-SPS}

\author{Christine Meurer (for the NA49 Collaboration)\dag\  
\footnote[3]{Christine.Meurer@cern.ch}
}

\address{\dag\ Gesellschaft f\"ur Schwerionenforschung, Planckstr. 1, D-64291 Darmstadt, Germany}

\begin{abstract}
New results on the production of $\Xi$ and $\Omega$ hyperons in Pb+Pb interactions at 40~$A$\hspace{0.08cm}GeV and $\Lambda$ at 30~$A$\hspace{0.08cm}GeV are presented. Transverse mass spectra as well as rapidity spectra of these hyperons are shown and compared to previously measured data at different beam energies. The energy dependence of hyperon production (4$\pi$ yields) is discussed. Additionally, the centrality dependence of $\Xi^{-}$ production at 40~$A$\hspace{0.08cm}GeV is presented.
\end{abstract}

\section{Introduction}

Measurements of strange particles provide information on the properties of strongly interacting matter at early stages of
%the hadronisation features of strongly interacting matter produced in 
heavy ion collisions. On the assumption that a deconfinement phase transition occurs in heavy ion collisions, we have to look for qualitative changes (onset phenomena) when varying external parameters like system size and energy density. Therefore, we study $\Lambda$, $\Xi$ and $\Omega$ production as a function of beam energy and $\Xi^{-}$ production dependent on the collision centrality. This contribution reports on the first results of the $\Lambda$ production at 30~$A$\hspace{0.08cm}GeV and $\Xi$ and $\Omega$ production at 40~$A$\hspace{0.08cm}GeV. 
% This results are compared to predictions of a hadron-gas model \cite{hg}.

\section{The NA49 experiment}

The NA49 experiment \cite{NA49setup} is a large acceptance hadron spectrometer which consists of four TPCs. The two vertex TPCs are located in a strong magnetic field. They allow the measurement of charge and momentum of the detected particles. The two main TPCs are outside the magnetic field. They are used for particle identification via dE/dx measurement. A veto calorimeter which is placed downstream the TPCs measures the energy of spectator nucleons from which the collision centrality can be inferred. 

\section{Transverse mass spectra and inverse slope parameters}

\begin{figure}[h]
\begin{center}
\includegraphics[scale=0.55]{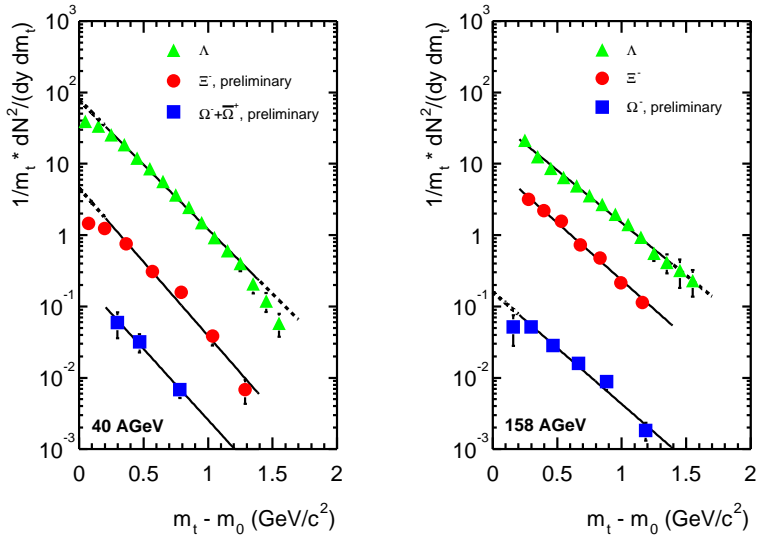}
\caption{\label{mt_spec} Transverse mass spectra of $\Lambda$ (triangles), $\Xi$ (circles), $\Omega$ (squares) produced in central Pb+Pb collisions at 40 (left) and 158~$A$\hspace{0.08cm}GeV (right). The $\Lambda$ spectra are uncorrected for feeddown from $\Xi$ and $\Omega$ decays.}
\end{center}
\end{figure}

In Fig.~\ref{mt_spec} results on $m_{T}$-spectra of hyperons at 40 and 158~$A$\hspace{0.08cm}GeV at midrapidity measured in central Pb+Pb collisions by NA49 are summarized. Because of the good signal to background ratio at 40~$A$\hspace{0.08cm}GeV $\Lambda$ and $\Xi$ cover almost in the full $m_{T}$-region. The $\Xi$ at 158~$A$\hspace{0.08cm}GeV have already been published \cite{rob}, while the $\Lambda$-spectra at both energies are submitted for publication \cite{andre}. All other presented results are still preliminary. 
%The $\Xi^{-}$ and $\Omega^{-}+\bar{\Omega}^{+}$-spectra at 40~$A$\hspace{0.08cm}GeV are presented here for the first time and they are still preliminary. 
At low and high $m_{T}$ the spectra deviate from an exponential shape. Because of this all spectra are fitted with a simple exponential function only in a limited $m_{T}-m_{0}$-range (0.2-1.4 GeV/$c^{2}$). From these fits the following inverse slope parameters at 40~$A$\hspace{0.08cm}GeV are derived: $T_{\Xi^{-}}=$ (210$\pm$11) MeV, $T_{\Omega^{-}+\bar{\Omega}^{+}}=$ (218$\pm$39) MeV. 
At 40~$A$\hspace{0.08cm}GeV the signals of $\Omega^{-}$ and $\bar{\Omega}^{+}$ hyperons are summed because of low statistics.
%Because of the low statistics of $\Omega^{-}$ and $\bar{\Omega}^{+}$ hyperons at 40~$A$\hspace{0.08cm}GeV the signals of particles and antiparticles are summed.

\section{Rapidity spectra and total yields}

\begin{figure}[h]
\begin{center}
\includegraphics[scale=0.55]{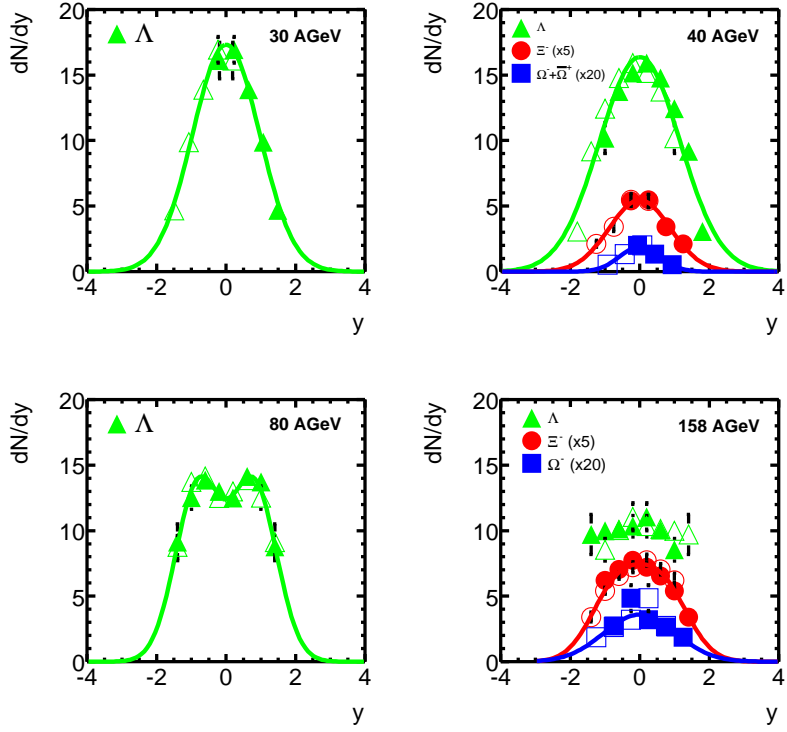}
\caption{\label{rap} Rapidity spectra of $\Lambda$ (triangles), $\Xi^{-}$ (circles), $\Omega$ (squares) produced in central Pb+Pb collisions at 30, 40, 80 and 158~$A$\hspace{0.08cm}GeV. Filled symbols are measured, open ones are reflected at midrapidity. For clarity the $\Xi$-spectra are scaled by a factor 5 and the $\Omega$-spectra by a factor 20. The $\Lambda$ spectra are uncorrected for feeddown from $\Xi$ and $\Omega$. All spectra are preliminary except $\Xi^{-}$ at 158~$A$\hspace{0.08cm}GeV.}
\end{center}
\end{figure}

The NA49 detector is able to measure particles in a broad rapidity range. In Fig.~\ref{rap} the rapidity distributions of $\Lambda$, $\Xi$ and $\Omega$ produced in central Pb+Pb collisions at several beam energies are shown. The $\Lambda$ rapidity spectra at 30~$A$\hspace{0.08cm}GeV, the $\Xi^{-}$ rapidity spectra at 40~$A$\hspace{0.08cm}GeV and the $\Omega$-spectra at 40 and 158~$A$\hspace{0.08cm}GeV are parametrised by Gaussians as indicated by the full lines. The $\Lambda$-spectrum at 80 and the $\Xi$-spectrum at 158~$A$\hspace{0.08cm}GeV are parametrised by the sum of two Gaussians displaced symmetrically with respect to midrapidity. A clear evolution of the shape of the $\Lambda$-spectra is visible. For $\Xi^{-}$ and $\Omega$ we observe an increase of the width of the rapidity distributions with beam energy, but no indication for a change of the shape.
The measured rapidity spectra have been integrated to obtain the total multiplicities in the full phase space, for numerical values see table~\ref{4piyield}.

\begin{table}
\caption{\label{4piyield} Total yields of the hyperons produced in central Pb+Pb collisions at several beam energies. The fraction of the selected cross section is given in the first row. The $\Lambda$ yields are uncorrected for feeddown from $\Xi$ and $\Omega$. The first given error is statistical and the second one systematic.}
\begin{tabular*}{\textwidth}{@{}l*{15}{@{\extracolsep{0pt plus12pt}}l}}
\br
beam energy & 30~$A$\hspace{0.08cm}GeV & 40~$A$\hspace{0.08cm}GeV & 80~$A$\hspace{0.08cm}GeV  & 158~$A$\hspace{0.08cm}GeV  \\
$\sqrt{s}$ & 7.62 GeV  &  8.73 GeV  & 12.30 GeV  & 17.30 GeV \\
\mr
$\sigma^{cent}/\sigma^{inel}$ & 7.2\% & 7.2\% & 7.2\% & 10\% ($\Omega$: 20\%) \\
$\langle \Lambda \rangle$ & 41.9$\pm$2.1$\pm$4.0 & 45.6$\pm$1.9$\pm$3.4 & 47.4$\pm$2.8$\pm$3.5 & 44.1$\pm$3.2$\pm$5.0 \\
$\langle \Xi \rangle$    & -- & 2.41$\pm$0.15$\pm$0.24 & -- & 4.12$\pm$0.20$\pm$0.62 \\
$\langle \Omega^{-}+\bar{\Omega}^{+} \rangle$ & -- & 0.20$\pm$0.03$\pm$0.04 & -- & 0.62$\pm$0.07\\
$\langle \Omega^{-} \rangle$ & -- & -- & -- &  0.47$\pm$0.07\\
\br
\end{tabular*}
\end{table}

\section{Energy dependence of hyperon production}

\begin{figure}[h]
\noindent
\begin{minipage}[t]{.49\linewidth}
%\begin{center}
\centering\epsfig{figure=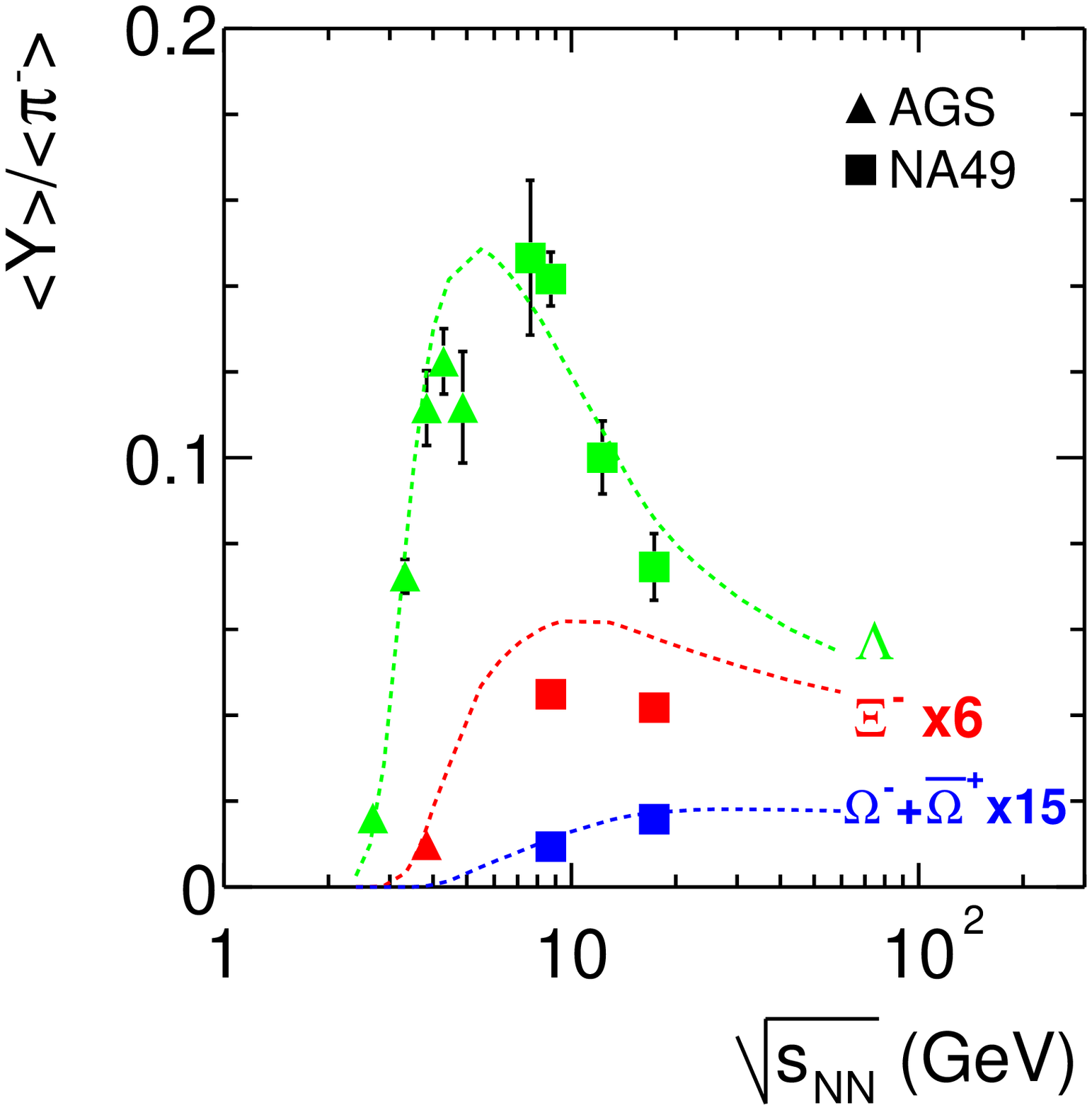,width=\linewidth,height=7.5cm}
\caption{\label{energy_dep} Energy dependence of the hyperon to $\pi^{-}$ ratio (measured in $4\pi$) compared to a prediction of the hadron-gas model. For clarity the $\Xi$ curve is scaled by a factor 6 and the $\Omega$ curve by a factor 15.}
\end{minipage}\hfill
\begin{minipage}[t]{.49\linewidth}
\centering\epsfig{figure=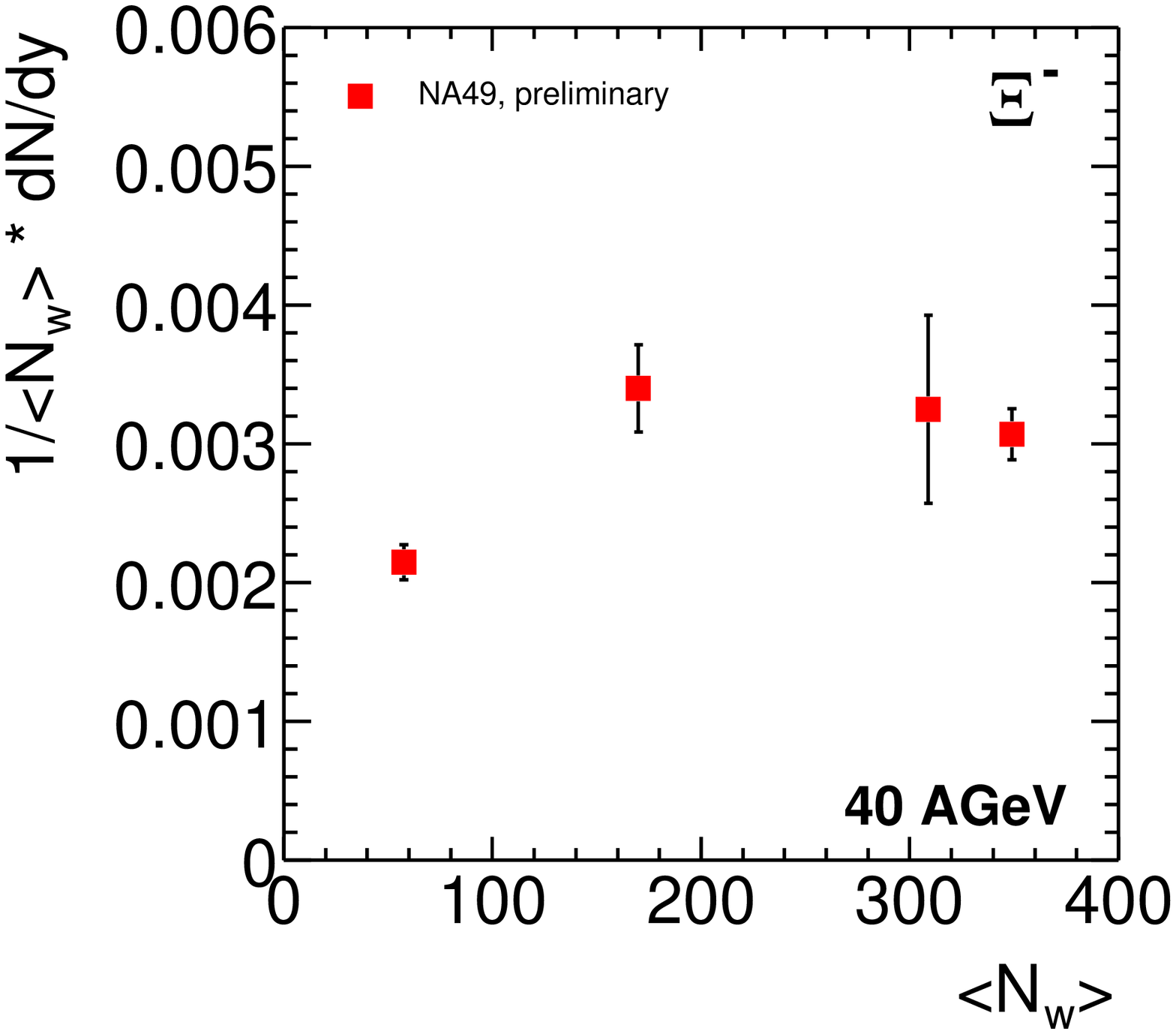,width=\linewidth,height=7.5cm}
\caption{\label{cen_dep} Centrality dependence of $\Xi^{-}$ production in Pb+Pb collisions at 40~$A$\hspace{0.08cm}GeV (midrapidity) as measured by NA49. 
%The squares shows the measurement of NA49 compared to a measurement of NA57 (circle). 
 All values are normalised by the mean number of wounded nucleons $N_{W}$.}
\end{minipage}
%\end{center}
\end{figure}

The total multiplicities of $\Lambda$, $\Xi$ and $\Omega$ are plotted as a function of the centre of mass energy in Fig.~\ref{energy_dep}. All yields are divided by the total multiplicities of the negatively charged pions. The energy dependence of the $\Lambda$ hyperons shows a clear maximum at about 30~$A$\hspace{0.08cm}GeV ($\sqrt{s}=$7.62 GeV). The $<\Xi^{-}>/<\pi^{-}>$ ratio shows a weak maximum at 40~$A$\hspace{0.08cm}GeV ($\sqrt{s}=$8.73 GeV). The non-monotonic behaviour is not seen for the $<\Omega^{-}+\bar{\Omega}^{+}>/<\pi^{-}>$ ratio.
The dashed lines show a prediction of a grand canonical hadron-gas model, which includes the variation of the baryo-chemical potential and the temperature with the beam energy \cite{hg}. The model describes the basic features of our data within 30\%.

\section{Centrality dependence of $\Xi^{-}$ production}

In Fig.~\ref{cen_dep} the centrality dependence of $\Xi^{-}$ production at 40~$A$\hspace{0.08cm}GeV at midrapidity is shown. The yields are normalised by the mean number of wounded nucleons \cite{glauber}.
The $\Xi^{-}$ production increases from peripheral to midcentral collisions and this behaviour is followed by a saturation toward central reactions. 
%For comparison a measurement of NA57 is included \cite{na57}.

\end{document}